# Stability of near surface nitrogen vacancy centers using dielectric surface passivation

*Ravi Kumar[1,*,‡], Saksham Mahajan[2,‡], Felix Donaldson[1], Siddharth Dhomkar[1,5,6], Hector J. Lancaster[4], Curran Kalha[3], Aysha A. Riaz[3], Yujiang Zhu,[3] Christopher A. Howard[4], Anna Regoutz[3] and John J.L. Morton[1,2]*

[1]London Centre for Nanotechnology, UCL, London WC1H 0AH, UK
[2]Department of Electronic & Electrical Engineering, UCL, London WC1E 7JE, UK
[3]Department of Chemistry, UCL, 20 Gordon Street, London WC1H 0AJ, UK
[4]Department of Physics and Astronomy, UCL, London WC1E 6BT, UK
[5]Department of Physics, IIT Madras, Chennai 600036, India
[6]Center for Quantum Information, Communication and Computing, IIT Madras, Chennai 600036, India

*Email: ucanrku@ucl.ac.uk
‡These authors contributed equally

Keywords: diamond, NV centers, quantum sensing, optically detected magnetic resonance

## Abstract

We study the photo-physical stability of ensemble near-surface nitrogen vacancy (NV) centers in diamond under vacuum and air. The optically detected magnetic resonance contrast of the NV centers was measured following exposure to laser illumination, showing opposing trends in air compared to vacuum (increasing by up to 9% and dropping by up to 25%, respectively). Characterization using Raman and X-ray photoelectron spectroscopies suggests a surface reconstruction: In air, atmospheric oxygen adsorption on surface leads to an increase in $NV^-$ fraction, whereas in vacuum, net oxygen desorption increases the $NV^0$ fraction. NV charge state switching is confirmed by photoluminescence spectroscopy. Deposition of ~ 2 nm alumina ($Al_2O_3$) over the diamond surface was shown to stabilize the NV charge state under illumination in either environment, attributed to a more stable surface electronegativity. The use of alumina coating on diamond is therefore a promising approach to improve the resilience of NV sensors.

**Introduction**

The nitrogen vacancy (NV) center in diamond is an atomistic defect which has emerged as a leading candidate in many solid state quantum technologies,[1,2] including quantum sensors to study diverse systems in fields ranging from solid state physics to complex biological environments.[3-5] The negative NV charge state (NV⁻) is central to such applications in quantum sensing owing to its optically addressable spin states, long-lived quantum coherence and room temperature operation.[6] The ground state spin-Hamiltonian of NV⁻ center is sensitive to various physical quantities which forms the basis of quantum sensing.[7-9] The performance of the NV-diamond sensor can be determined by measurement sensitivity ($\eta_{mag.} \propto \sqrt{1 + \frac{1}{C^2 n_{avg.}}}$ ; Where $\eta_{mag.}$ is AC or DC magnetic sensitivity, C is measurement contrast and $n_{avg.}$ is number of NV⁻ photons per measurement)[10] in addition to the spatial resolution (as low as ~ 10 nm)[11] and operational stability under various environments. Achieving the greatest spatial resolution for quantum sensing requires the positioning of NV centers near diamond surface (< 10 nm), however, the brightness and spin properties of NV centers are compromised near the surface.[2] For example, the DC magnetic field sensitivity ($\eta_{mag.}$) achieved is ~ 17 $pT/\sqrt{Hz}$ for NV centers in the bulk,[12] which can be compared to ~ 1 $\mu T/\sqrt{Hz}$ for near-surface NV centers.[13] Furthermore, the stability of near surface NV centers under non-ambient conditions is critical for studying various temperature and pressure dependent physical phenomenon in solids like magnetic and superconducting materials.[14-16]

The most general diamond surface composition involves non-diamond ($sp^2$) carbon, functional groups ($C-X_n$,), dangling bonds, metallic traces and adsorbed environmental species.[17,18] Among these, many surface constituents have been identified as a source of local charge traps (e.g. $sp^2$ carbon is known to form the double potential well as electronic trap state)[19] leading to fluctuating electric and magnetic fields which would degrade the properties of proximal NV centers.[20-22] NV-diamond quantum sensing protocols typically use high power non-resonant laser excitation (~ 532 nm)[12], which can lead to significant heating, NV spectral diffusion, ionization of nitrogen atoms (P1 centers) and excitation of various surface constituents [23-25]. To improve the stability of near surface NV centers under ambient conditions, numerous atomic functionalization and organic species have been applied on the surface.[19, 26-30] Ultra-high vacuum (UHV) and cryogenic conditions have been reported to degrade the properties of single NV centers in diamond nano-pillar structures, while their stability was partially improved upon surface passivation with ultra-pure water.[31] The exact origin of NV degradation near surface and under different environmental conditions remains unclear and requires attention in order to develop effective mitigation strategies.

In this article, we investigate the instability of near surface ensemble NV centers under air and vacuum (~1 ×10$^{-3}$ mbar) at room temperature, measuring NV properties such as optically detected magnetic resonance (ODMR) contrast as well as characteristic properties of the material surface, following laser illumination. We found the ODMR contrast to vary over the course of (1-2 mW) laser exposure, due to NV charge state conversion (NV$^-$ ↔ NV$^0$) caused by the changing surface chemistry. A stable ODMR contrast and NV charge state were achieved by atomic layer deposition (ALD) of aluminium oxide (Al$_2$O$_3$, ~ 2 nm) on the diamond surface.

**Experimental details:**

The primary sample investigated here is an electronic grade (100) diamond, ion implanted with nitrogen ($^{15}$N, 3 keV, 1×10$^{13}$ ions/cm$^2$), supplied by Qnami AG.[32] The average NV depth was estimated by average range of N$^+$ ions and lattice vacancy profile using stopping and range of ions in matter (SRIM) to be ~ 5 nm (Fig. S1; supporting information (SI)). The concentration of NV centers was ~ 5000 NVs/μm$^2$. The as-procured sample was acid refluxed at 255 °C (H$_2$SO$_4$: HClO$_4$: HNO$_3$ with 1:1:1 v/v ratio) for 2 hours to eliminate non-diamond impurities and increase the oxygen functionalization. We used the acid reflux as a procedure to 'reset' the diamond surface in between experiments in different environments. The acid refluxed sample was termed NV-diamond ('NVD'). Sample preparation has been summarized in fig. 1(b). After the completion of optical measurements on NVD, a ~ 2 nm layer of aluminum oxide (Al$_2$O$_3$) layer was deposited on the sample (NVD→Acid reflux →Al$_2$O$_3$) using a Savannah S200 atomic layer deposition (ALD) system. The thickness of the Al$_2$O$_3$ layer was found to be ~ 2 nm, measured by ellipsometry on a bare silicon substrate (placed together with NVD during ALD deposition). The sample with the deposited Al$_2$O$_3$ layer is termed alumina coated NV-diamond ('AC-NVD').

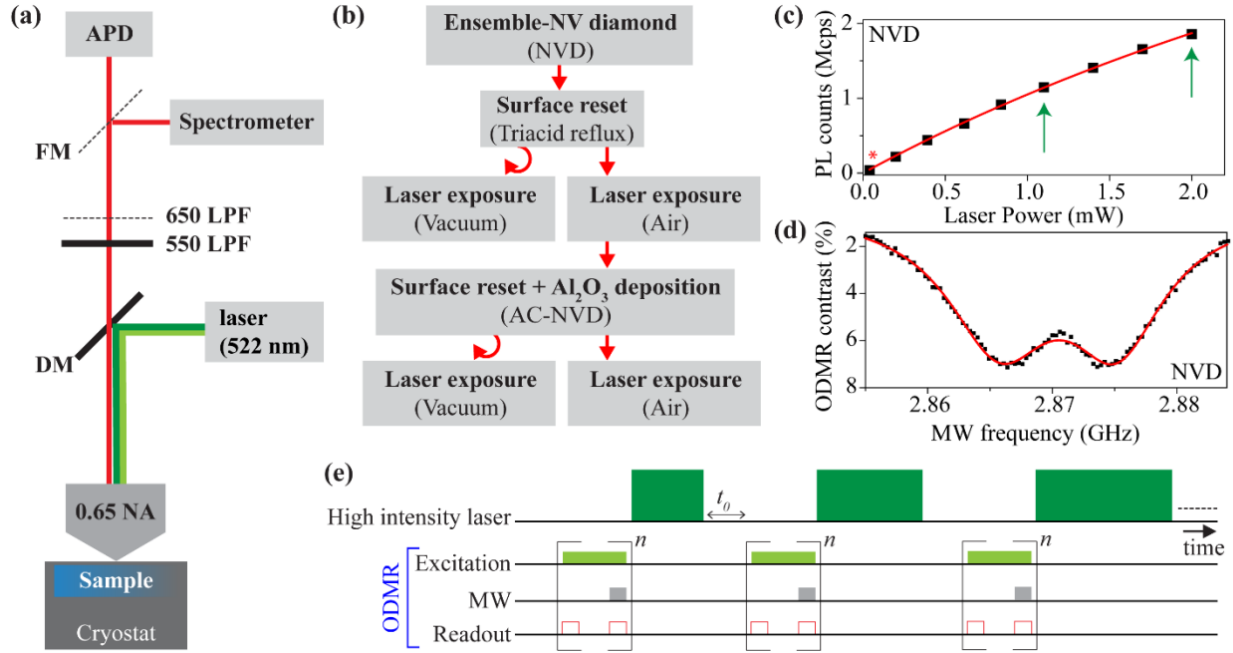

**Figure 1**. (a) Schematic of optical measurement setup used for PL and ODMR. (b) Summary of the sample preparation and optical measurement steps. (c) Fluorescence saturation measurements (squares) for NVD sample with a fit (curve) to the saturation curve (see text). Red shaded regions show the PL signal counts at powers used in subsequent laser exposure experiments. (d) Continuous wave (CW) ODMR spectrum for NVD sample (black dots) and double Lorentzian fitting (solid red line). (e) Pulse sequence to study ODMR contrast under the influence of periodic high power laser pulses of increasing duration.

The experimental setup for optical measurements (see Fig. 1(a)) consists of a home built confocal setup (NA = 0.65) equipped with a Montana instruments s100 cryostation and 522 nm continuous wave laser (LBX-522; Oxxius). The fluorescence signal was filtered through flip mounted 550 nm and 650 nm long pass filters (LPFs) and guided toward a single photon counting module (Excelitas Technologies) and photoluminescence (PL) spectrometer (SpectraPro HRS500, Princeton instruments) for ODMR and PL spectroscopy respectively. For ODMR, the fluorescence signal was collected through a 550 nm LPF in order to observe maximum effect of $NV^0$ emission in ODMR measurements. A laser power of 40 µW was used for ODMR and PL spectroscopy measurements. To study the impact on ODMR contrast from high power laser illumination, ~ 1.2 mW and ~ 2 mW laser powers were used. Figure 1(c) shows the PL intensity as a function of laser power $P$, fitted to the function: $\frac{P \cdot I_{sat.}}{(P+P_{sat.})} + \alpha P$ where $I_{sat.}$ is the saturated intensity (8.5 ± 0.5) Mcps, $P_{sat.}$ the saturation power (7.1 ± 0.6) mW and $\alpha$ denotes the (non-saturating) background fluorescence. The PL counts shown in Fig. 1(c) as a function of laser power were measured after inserting a neutral density (ND) filter in optical collection path – the actual counts are expected to be ~10 times higher. For vacuum measurements, the sample chamber was evacuated using a rotary pump ($1\times10^{-3}$ mbar), while for measurements under ambient conditions the cryostat head was removed. The experimental scheme for ODMR measurements is shown in Fig. 1(e). High power laser illumination was repeatedly applied to the sample with increasing illumination time after each repetition. The ODMR spectrum was measured using lower powers ($P_{exc.}$ = 40 µW) after each high power exposure, following a wait time ($t_0$) of 60 s. After a cumulative exposure time of ~ 1.5 $\times 10^4$ s at ~ 1.2 mW, laser power was increased to ~ 2 mW up to a total illumination time of ~ 3.0 $\times 10^4$ s.

Additional material characterization was performed using high purity electronic grade diamond (ELSC20, Thorlabs). As received ELSC20 samples were acid refluxed and termed as electronic grade diamond (ED). A ~ 2 nm layer of $Al_2O_3$ was deposited on ED samples and termed as alumina coated electronic grade diamond (AC-ED). In-situ Raman spectroscopy was performed using Renishaw in-Via Raman microscope equipped with 514.5 nm laser. To estimate the effect of laser illumination, spectral features between ~ 1200 – 1900 $cm^{-1}$ were recorded repeatedly under continuous high laser power excitation (2 mW power was applied through an air objective lens of 0.4 NA). For vacuum Raman measurements, the ED and AC-ED samples were placed in a custom made vacuum compatible glass cell and evacuated to ~ $1\times10^{-5}$ mbar. The sealed glass cell was then placed under the microscope for spectroscopy. Other Raman measurements were performed under ambient conditions. Normalized Raman spectra were deconvoluted using Lorentzian line shapes to compare diamond (~ 1332 $cm^{-1}$) and graphitic carbon (~ 1600 $cm^{-1}$; G band) signal intensities. For XPS sample preparation, the 2D Raman imaging (Area ~ 125 µm$^2$) of EDs was performed under different environmental conditions. Two laser exposed samples were prepared under air and vacuum environmental conditions respectively (Figure S5; SI) for XPS. To eliminate adventitious carbon and adsorbed surface species, the laser exposed EDs were annealed at 200 ºC (2 hours) under argon atmosphere prior to XPS. The XPS was performed using a Thermo Scientific K-Alpha X-ray photoelectron spectrometer with a base pressure of ~ $2 \times 10^{-9}$ mbar, equipped with a monochromatic Al $K_\alpha$ X-ray source (hv = 1486.7 eV). The X-ray spot size was reduced from the standard 400 µm to 100 µm in order to resolve the laser exposed regions in both samples. The XPS spectra for each sample were recorded at laser exposed position and another unexposed position (situated at ~ 1 mm away from laser exposed position). The maximum XPS probing depth ($d_{XPS}$) at the maximum kinetic energy of 1486.7 eV, i.e. the photon energy of the Al Kα laboratory X-ray source, was estimated by calculating the relativistic inelastic mean free path (IMFP) ($d_{XPS} = 3 \times (IMFP)$) using the TPP-2M model as implemented in the QUASES software package.[33] The $d_{XPS}$ for the diamond and diamond with $Al_2O_3$ samples were calculated based on C and $Al_2O_3$ models available in the QUASES database and were found to be ~ 11.7 nm and 10.2 nm, respectively. XPS analysis was performed using the Thermo Avantage software package. For the estimation of the relative atomic ratios of carbon and oxygen in different samples, the total peak areas of the C 1$s$ and O 1$s$ core levels and in-built atomic sensitivity factors (ASFs) were used. The change in carbon to oxygen atomic ratio (C/O) due to laser exposure was quantified as $\Delta(C/O) = \frac{(C/O)_{exp.} - (C/O)_{unexp.}}{(C/O)_{unexp.}} \times 100$, where $(C/O)_{unexp.}$ and $(C/O)_{exp.}$ represent unexposed and laser exposed positions, respectively. Core level spectra were fitted using the smart background function and Lorentzian-Gaussian sum functions (Figs. S6 & S7). The graphitic ($sp^2$) and diamond ($sp^3$) carbon peak contributions were extracted and the $sp^2/sp^3$ ratio derived. The change in $sp^2/sp^3$ ratio due to laser exposure was quantified as $\Delta\left(sp^2/sp^3\right) = \frac{(sp^2/sp^3)_{exp.} - (sp^2/sp^3)_{unexp.}}{(sp^2/sp^3)_{unexp.}} \times 100$; where $(sp^2/sp^3)_{exp.}$ and $(sp^2/sp^3)_{unexp.}$ represent the laser exposed and unexposed positions on the sample.

**Results and discussion:**

The spin state dependent brightness of the NV$^-$ center is a fundamental part of its application as a quantum sensor and can be characterised by the ODMR contrast, or the relative change in PL intensity following a change in the spin state.[6] The ground state ($^3$A) spin triplet ($m_s = 0, \pm 1$) of the defect is characterized by an axial zero field splitting (ZFS) of about 2.87 GHz between spin sublevels $m_s = 0$ and $m_s = \pm 1$. Due to non-axial lattice strain induced by P1 centers, unoccupied vacancies, implantation induced lattice structural defects and local electric field, the degeneracy of spin sub-levels $m_s = \pm 1$ is also lifted by a non-axial ZFS which can vary from 100 kHz to a few MHz.[34-36] Therefore, in the absence of an external magnetic field, the ODMR spectrum is characterized by a resonance around 2.87 GHz, further split by the non-axial term, as illustrated by the representative ODMR spectrum for NVD in an air environment shown in Fig. 1(d). The maximum ODMR contrast for a single NV is about 30%. For ensemble NVs, the ODMR contrast reduces significantly due to several factors such as strain induced line broadening, interactions with paramagnetic impurities, non-trivial charge dynamics and inefficient pi-pulse for different NV orientations. [37-42]. The evolution of ODMR contrast under laser illumination in different surface and environmental conditions is shown in Fig. 2. For the NVD sample under vacuum, the ODMR contrast exhibits an exponential decay with laser exposure at 1.2 mW, and decays further when the laser power increased to 2 mW (Fig. 2(a)), dropping to a contrast of ~4.5% (or ×0.75 the starting value). The opposite trend is seen when illuminating NVD in air (Fig. 2(b)), where the contrast is seen to rise to ~7.8% (or ×1.09 the starting value). However, for the AC-NVD sample, the ODMR contrast was found to be relatively independent on high power laser illumination under either environments, changing by less than ~ 0.3%. We fit the time evolution of ODMR contrast under the two consecutive periods of laser exposure at different powers, using two exponential functions with a common set of fitting parameters to ensure the evolution of the contrast is continuous across the two periods. Specifically, we use the functions $y(t) = C_0 + (C_1 e^{-t/\tau_1} + C_2)$ for t < 14732 s and $y(t) = C_0 + (C_1 e^{-14731/\tau_1} + C_2 e^{-(t-14731)/\tau_2})$. Here, $\tau_1$ and $\tau_2$ are the decay constants under laser exposure of 1.2 mW and 2 mW respectively. $C_0$ is the ODMR contrast after high laser illumination for infinite time. The $C_1$, and $C_2$ denotes the change in the ODMR contrast after laser illumination of 1.2 mW and 2 mW for infinite time.

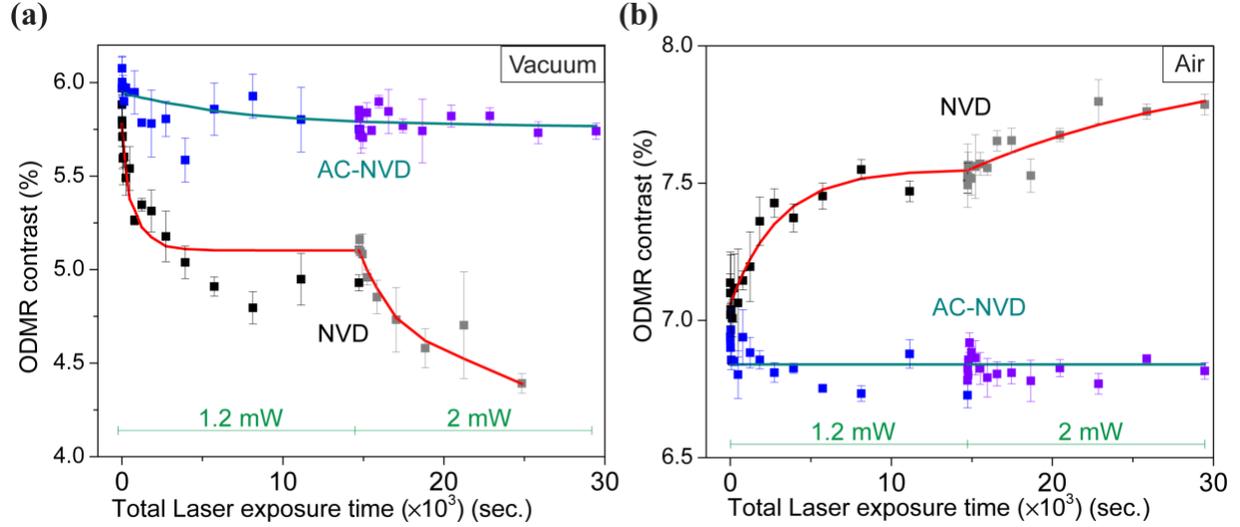

Figure 2. Evolution of ODMR contrast as a function of the total duration of laser power exposure for NVD and AC-NVD samples under (a) vacuum and (b) air environments. A series of laser exposures of increasing duration (from 1 s to 3.6 ks) is first applied using 1.2 mW laser power. After a cumulative exposure time of about 15 ks, the laser power is increased to 2 mW, and the exposure time per datapoint reset to 1 s time and subsequently increased. The data are fit to exponential decay functions, with separate time constants for the periods of 1.2 mW and 2 mW laser exposure (see text).

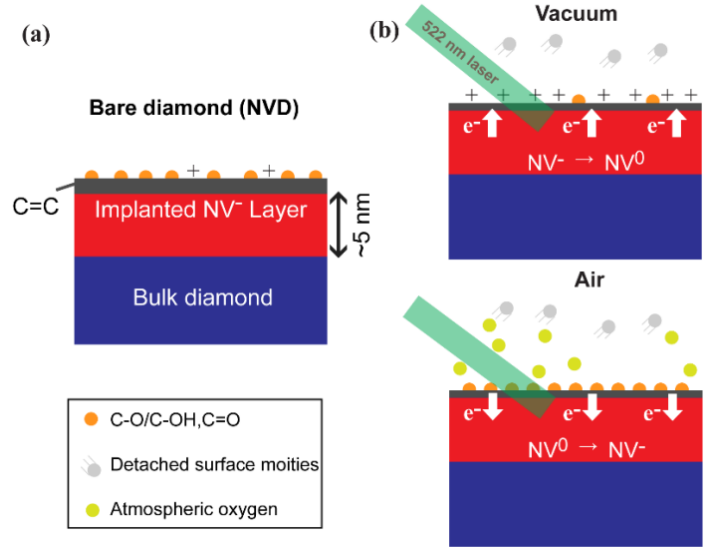

**Figure 3**. Schematic for proposed mechanism (a) The acid refluxed diamond sample (NVD) has a shallow NV-doped layer up to 5 nm from the surface, which consists of non-diamond carbon (grey region) and oxygen functionalities (orange dots) on the surface. (b) Laser exposure under different environments can cause desorption of non-diamond carbon and oxygen-containing functional groups (in vacuum), or an increase in oxygen termination on the surface (in air). These changes lead to charge state conversion between NV⁻ and NV⁰.

We attribute the observed changes in ODMR contrast to the conversion of NV⁻ to NV⁰, through a mechanism illustrated in Fig. 3, supported by measurements described in Figs. 4 and 5. Due to near surface NV fabrication (3 keV N⁺ ions), a high N/NV ratio (~ 1%) is present in the NVD sample,[43] and the residual nitrogen atoms (P1 centers) act as a source of electrons to maintain NV⁻ as preferential NV charge state.[44, 45] The high electronegativity of the oxygen functionalized surface also helps to maintain their stability.[17] During laser exposure, excitation of surface species[46] can result in the detachment of non-diamond carbon and oxygen functionalities and, in the absence of environmental oxygen (e.g. in vacuum), surface electron traps develop which reduce the surface electronegativity, cause upward band bending near the surface and lead to NV charge state conversion. The reduction in NV⁻ PL emission on top of a background fluorescence signal leads to a gradual reduction in the observed ODMR contrast, as a result of this continuously evolving surface composition. Within an oxygen-rich environment (e.g. in air), laser illumination has the same impact on reducing non-diamond carbon at the surface, however, there is an increase in the oxygen adsorption which increases surface electronegativity (downward band bending) and promotes the NV$^-$ charge state (NV$^0$ →NV$^-$). The evolution of the surface described above (and associated NV charge state conversion) appears to be minimised by the presence of the alumina coating in the AC-NVD sample. Any changes at the diamond surface may be compensated by the alumina providing a stable surface electronegativity on surface.

To further investigate the NV center charge state dynamics we monitored the PL features of NV$^0$ and NV$^-$, which are respectively characterised by zero phonon lines (ZPLs) at 575 nm and 637 nm accompanied by broad sideband emission with maxima around 640 nm and 700 nm.[7] Conversion from NV$^-$ to NV$^0$ leads to a relative increase (decrease) in PL intensity below (above) ~700 nm, as illustrated in Fig. 4(a). We recorded PL spectra using ~ 40 μW laser power, before and after 1.2 mW laser exposure for one hour. The difference PL spectra obtained by normalizing and subtracting the spectra before and after high power laser exposure are shown in Fig. 4(b) and (c). Normalized PL spectra under different experimental conditions are shown in Fig. S3. The PL spectrum for NVD under vacuum confirms the increase in $NV^0$ emission intensity (in the range 550 – 670 nm) and decrease in $NV^-$ intensity (670 – 800 nm), and the opposite behaviour is observed in an air environment. In both cases, negligible changes in the PL spectrum are seen for the AC-NVD sample under vacuum, consistent with NV charge state stability. The comparison of normalized PL of NVD and AC-NVD samples under different environments confirmed that alumina layer itself does not induce background fluorescence (Fig. S4, SI)). Overall, these results are consistent with the observed changes in ODMR contrast and mechanism described in Figure 3.

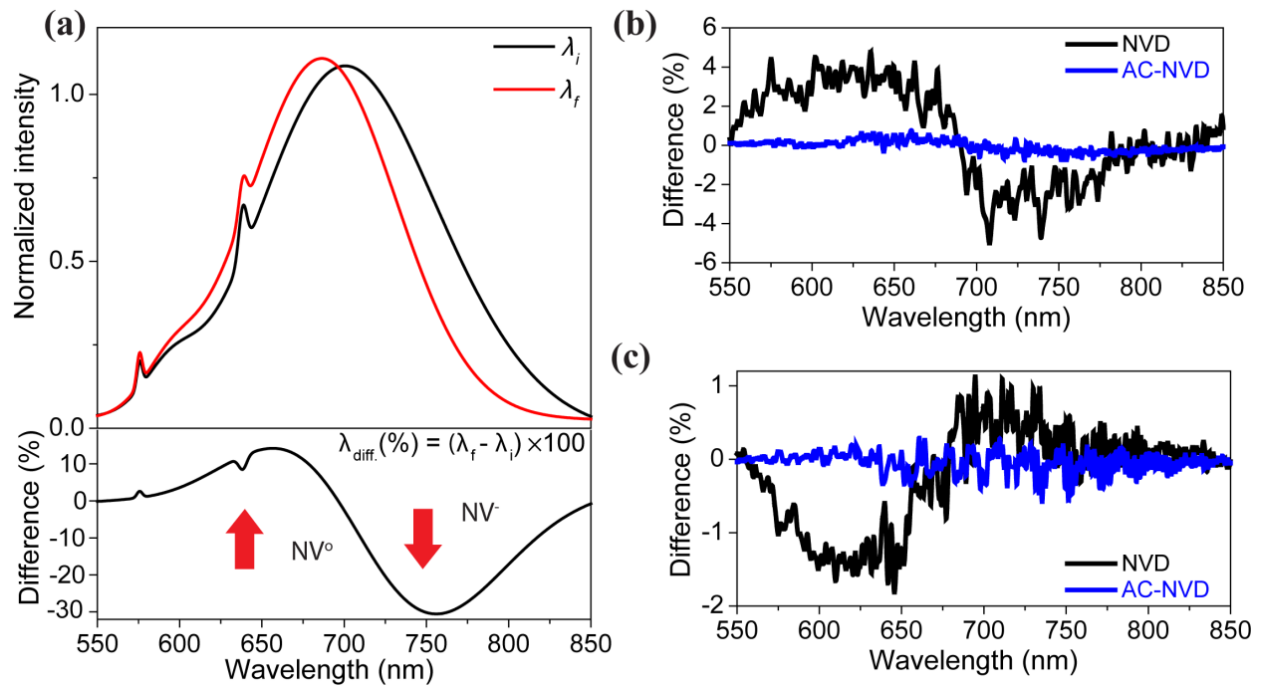

**Figure 4**. (a) Cartoon representation showing how the PL spectrum evolves as the NV$^-$/NV$^0$ fraction changes, with the difference shown in the lower panel. Measured difference PL spectra for NVD and AC-NVD are shown (b) under vacuum and (c) in air.

To gain further insights into material origin of observed ODMR contrast changes, XPS and in-situ Raman spectroscopy were performed (see Fig. 5). The surface composition dynamics due to high laser power exposure under different environmental conditions were analyzed by XPS (for details of sample preparation, see Fig. S5).[30] The ED sample exhibits only carbon (C 1$s$) and oxygen (O 1$s$) elements within the XPS probing depth (< 11.7 nm calculated based on the maximum IMFP). For AC-ED, the $Al_2O_3$ layer was also observed (evidenced by characteristic aluminium peaks (Al 2$p$ and Al 2$s$) and an intense O 1$s$ peak) in addition to carbon and oxygen (Fig. 5(a)).[47] The carbon to oxygen relative atomic ratios (C/O) reveal significant surface reconstruction in ED due to laser exposure (Fig. S6). The change in the C/O ratio comparing exposed and unexposed positions indicates that laser exposure in vacuum results in more efficient surface oxygen detachment compared to that in air (Fig. 5(b)). The C 1$s$ core level spectra for different samples were calibrated (peak fitting was used to determine the $sp^3$ peak position) to the reported binding energy value for diamond (285.0 eV) and are shown in Fig. 5(c).[26, 48] The C 1$s$ core level spectra of ED when laser exposed in different environments show a variation in spectral shape because of laser exposure (Fig. 5(c)). Peak fitting of the C 1$s$ core level spectra was performed to disentangle, identify, and quantify the different carbon-related chemical states (Fig. S6). In ED, non-diamond carbon ($sp^2$) is found to be present in addition to diamond ($sp^3$) (SI, Fig. S6(c)).[49, 50] The reduction in the ratio of $sp^2$ to $sp^3$ under laser exposure is much greater in air than in vacuum (Fig. 5(d)). Peak fitting of the O 1$s$ core level spectra (Fig. S7) reveals that the rate of elimination of specific oxygen functionalities (ether/alcohol (C-O-C/C-O-H) or ketone (C=O)) depends on the environment: Due to efficient etching of $sp^2$ carbon during laser exposure under air, the concentration of C-O bonds increases whereas that of C=O bonds decreases. Laser exposure under vacuum induces less

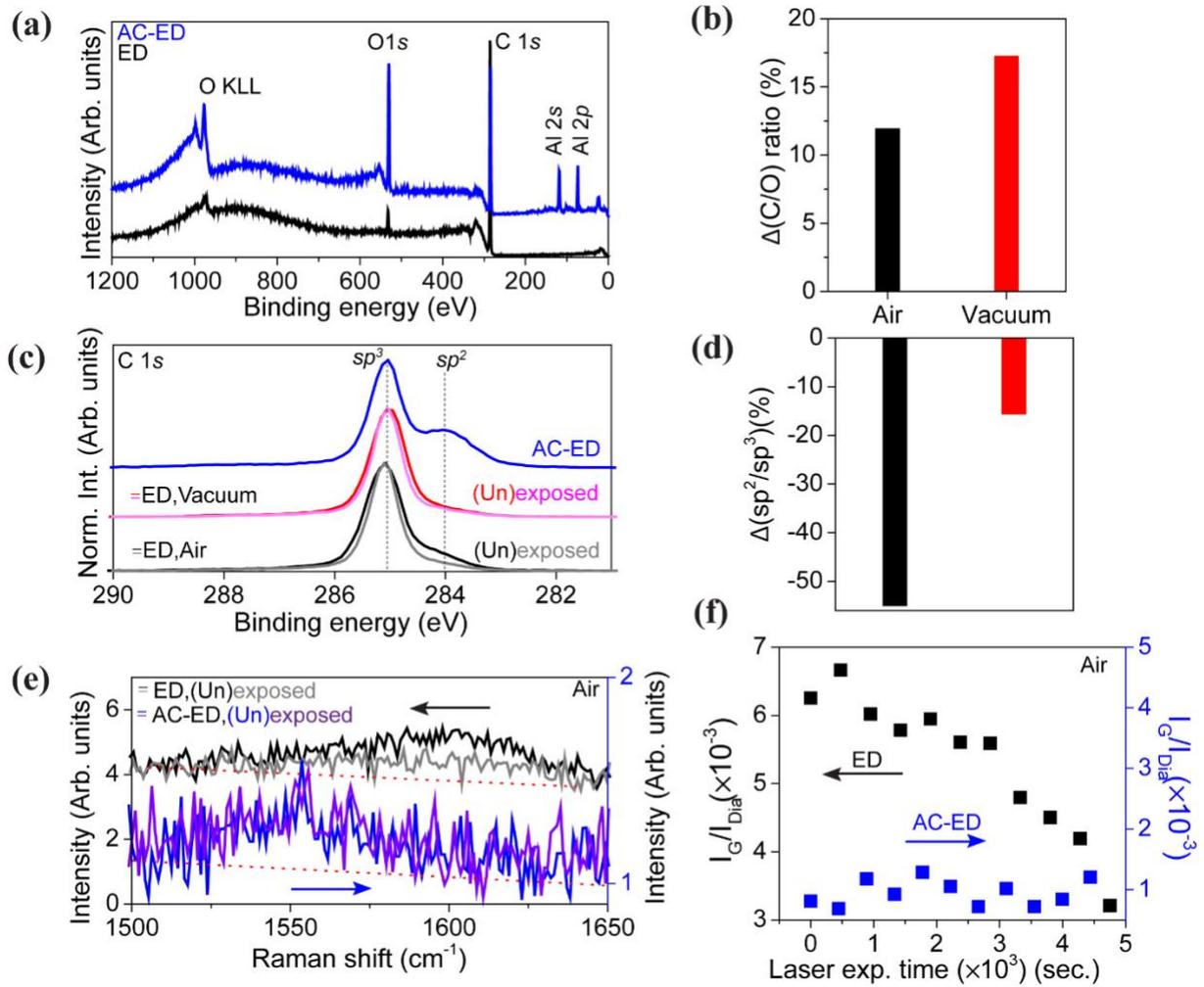

**Figure 5**. Surface spectroscopy (a) XPS survey spectra of ED and AC-ED samples before laser exposure. (b) Changes in the total carbon to oxygen (C/O) ratio of the ED sample as a result of laser exposure, in different environments. (c) C 1$s$ core level spectra showing results from regions in ED exposed to laser (pink, grey), as well as unexposed regions (black, red). The spectra are normalized to the maximum peak height. (d) Changes in the $sp^2/sp^3$ carbon ratio of the ED sample as a result of laser exposure, in different environments. (e) Normalised Raman spectra for ED and AC-ED samples under high power laser illumination in air. Red dotted lines are the baselines used for estimating G band intensity for ED and AC-ED. (f) The change in graphitic to diamond carbon ratio ($I_G/I_{Dia}$) as a result of high-power laser exposure under air environment.

efficient $sp^2$ carbon etching and therefore the rates of removal of C-O and C=O remain similar. In summary, these XPS measurements are consistent with the proposed mechanism (see Fig. 3), that surface oxygen is detached as a result of laser illumination. The C 1$s$ core level spectrum for AC-

ED (Fig. 5(c)) reveals a more significant fraction of $sp^2$ carbon compared to $sp^3$. This can be explained by a change in signal intensity from the diamond sample itself when the ~ 2 nm $Al_2O_3$ layer is added on top. This leads to a relative increase in the signal seen from the Diamond surface ($sp^2$) compared to its bulk ($sp^3$). In addition, the O 1$s$ core level spectrum of AC-ED is dominated by $Al_2O_3$, making it difficult to evaluate the diamond surface oxygen functionalization. Due to these factors, we did not perform XPS measurements on laser exposed AC-ED sample. However, it will be interesting to explore the AC-ED surface further to see if $Al_2O_3$ alters the diamond functionalization and how the diamond functionality varies under laser exposure.

The material changes due to laser exposure were further studied by Raman spectroscopy (Fig. 5(e,f)). Raman spectra of ED and AC-ED samples acquired under low laser excitation power demonstrated the absence of non-diamond carbon and related defects (Fig. S8) in addition to a sharp peak at ~1331.8 cm$^{-1}$ characteristic of diamond. To observe the effects of laser exposure, in-situ Raman measurements were performed (see Fig. S8 and SI for details). ED showed detectable G band features (centered at 1590 cm$^{-1}$), indicative of graphitic carbon, whose intensity reduced significantly after $4.8 \times 10^3$ s of continuous high power laser exposure. For AC-ED sample, a G band feature was observed at ~1550 cm$^{-1}$ (blue shifted compared to ED) and remained unchanged under prolonged laser exposure. The diamond related Raman features remained unchanged for both samples during high laser power exposure (Fig. S8). These observations are in agreement with XPS and suggest etching of $sp^2$ carbon in ED during laser exposure in air. We could not find detectable G band for either the ED or AC-ED sample under laser exposure in vacuum, possibly due to significant reduction in laser excitation intensity by vacuum glass cell.

**Summary and conclusion:**

The instability of near surface NV centers under non-ambient conditions is a long-standing challenge for diamond-based quantum sensing, particularly for work at cryogenic temperatures. To understand the origin of this instability, we have combined a study of the optical properties of the NV centers with an analysis of the material composition of the diamond surface. We observe a change of the ODMR contrast for near-surface (~ 5 nm deep) NV centers under laser illumination in oxygen functionalized diamond which we attribute to surface reconstruction under different environmental conditions. In vacuum, the ODMR contrast was reduced from around 6% to below

4.5%, whereas it increased under air up to over 7.5%. A ~ 2 nm layer of $Al_2O_3$ was deposited on diamond surface which successfully led to a stable ODMR contrast, even under laser exposure. The origin behind these changes in ODMR contrast was revealed by PL, XPS and Raman spectroscopies to arise from NV charge state switching caused by surface dynamics. In vacuum, owing to lack of atmospheric gases, electron traps develop on surface and the $NV^-$ charge state is converted into $NV^0$. Atmospheric oxygen inhibits the development of such traps and increases the $NV^-$ charge state fraction. The $Al_2O_3$ layer prohibits both the degradation of surface as well as adsorption of environmental oxygen, achieving a more stable NV charge state. The $Al_2O_3$-oxygen-diamond surface is shown to be resilient against optical excitation in vacuum, but requires further investigation under low temperature conditions, while the NV spin coherence properties should be analyzed in such material to assess its potential for quantum sensing. The use of alumina coating could be extended to help stabilise single near surface NV centers in planar and nanopillar diamond structures. A single NV-diamond AFM probe with stability under different environmental conditions might be achievable using such optimized passivation of the surface.

**Corresponding Author**

*Ravi Kumar: London Centre for Nanotechnology, UCL, London WC1H 0AH, UK
Email: ucanrku@ucl.ac.uk

**Author Contribution**

The manuscript was written through contributions of all authors. All authors have given approval to the final version of the manuscript. Ravi Kumar and Saksham Mahajan contributed equally.

**Acknowledgement**

This research has received funding from the Engineering and Physical Sciences Research Council (EPSRC) via the Centre for Doctoral Training in Delivering Quantum Technologies (EP/L015242/1) and the Hub in Quantum Computing and Simulation (EP/T001062/1), as well as from the European Research Council (ERC) via the LOQO-MOTIONS grant (H2020- EU.1.1., Grant No. 771493). We thank Dr. Ania C. Bleszynski Jayich (Department of Physics, University of California, Santa Barbara, USA), Dr. Felipe Fávaro de Oliveira (Qnami AG, Switzerland) & Gediminas Seniutinas (Qnami AG, Switzerland) for useful discussion. We thank Aferdita

Xhameni (Department of Electronic & Electrical Engineering, UCL) and Patrick Hogan (Department of Electronic & Electrical Engineering, UCL) for their help during the experiments. C.K. and A.A.R acknowledge the support from the Department of Chemistry, UCL.

**Supporting Information**

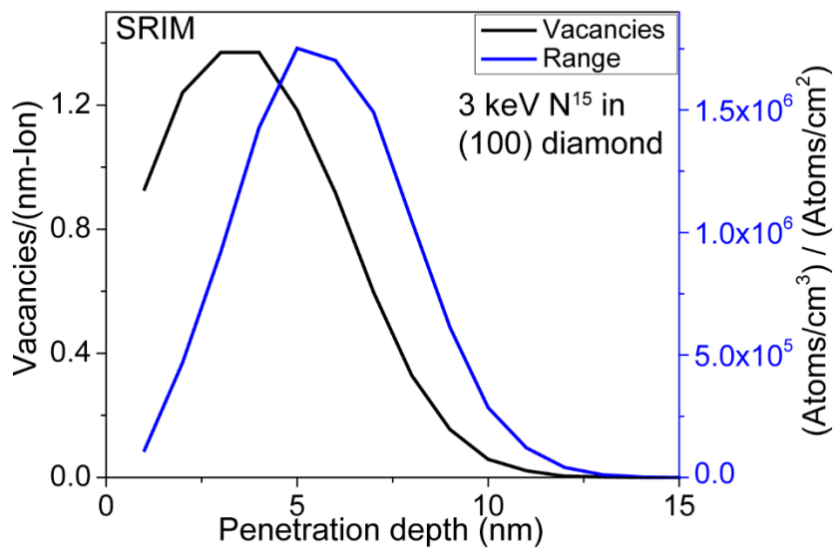

**Figure S1**. SRIM simulation of 3 keV $^{15}$N implantation in (100) diamond. The threshold energy ($T_d$) for displacement of carbon atoms from lattice position was taken to be 37.5 eV.[51] The average depth of lattice vacancies was ~ 3.5 nm (maximum depth of NV ~ 12 nm). Whereas, the average penetration depth was ~ 5.5 nm (maximum depth of implanted Nitrogen ~ 15 nm). Considering that nitrogen ions remain immobile during annealing procedure (post implantation NV fabrication step), the NV concentration profile should closely follow nitrogen stopping range curve (Blue).

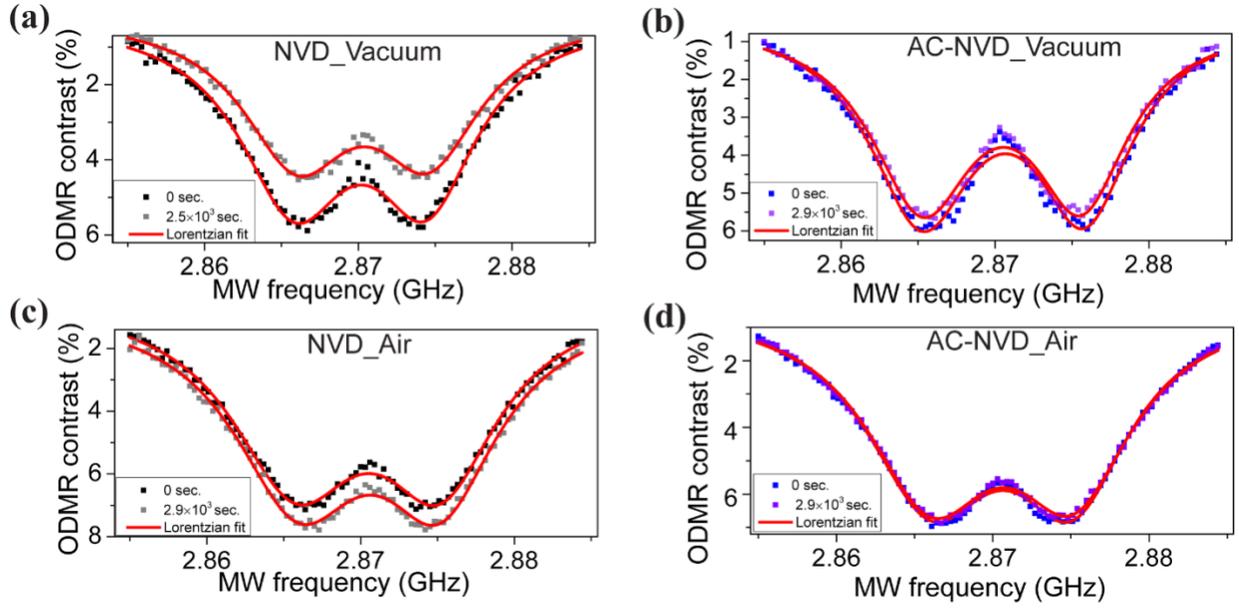

**Figure S2**. ODMR spectra before and after laser exposure for (a) NVD under vacuum (b) AC-NVD under vacuum (c) NVD under air and (d) AC-NVD under air respectively.

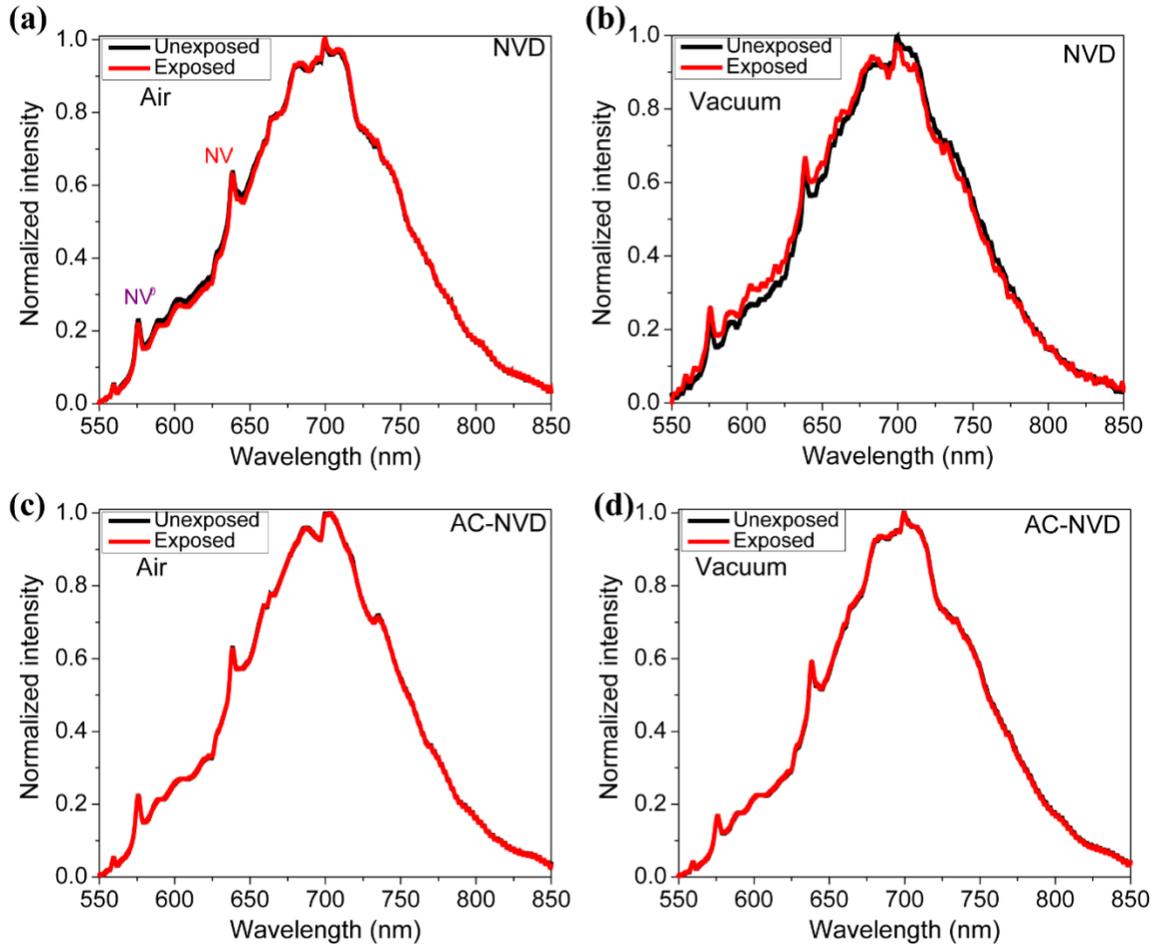

**Figure S3**. Normalized PL spectra under different conditions before and after high power laser exposure. (a) and (b) show normalized spectra for BD sample under air and vacuum respectively. (c) and (d) show the PL spectra of AC-ED sample under air and vacuum respectively.

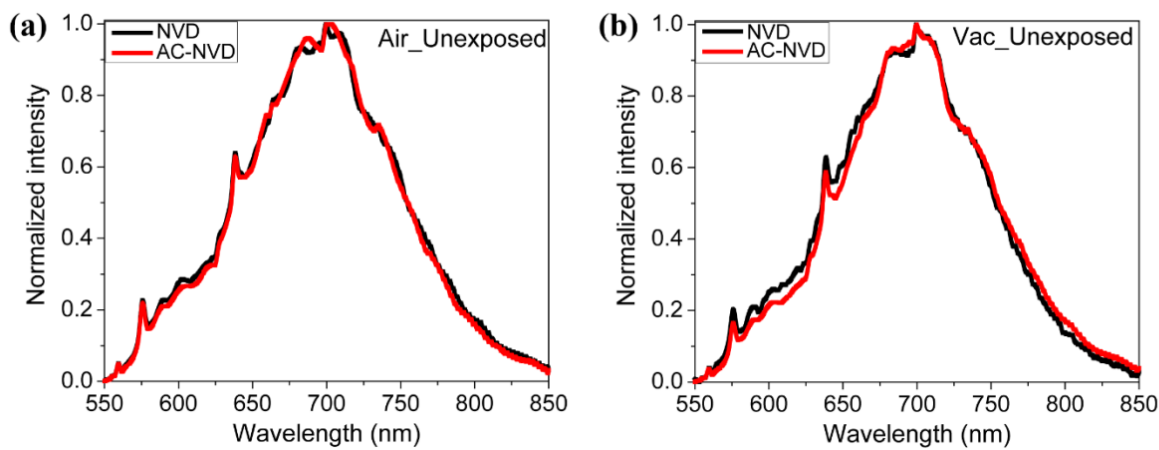

**Figure S4**. Normalized PL spectra of NVD and AC-NVD samples before high power laser exposure under (a) air and (b) vacuum respectively.

The sample preparation for the XPS has been summarized in fig. S5. Fig. S6(a) shows the Al $2p$ core level spectrum for AC-ED. Fig. S6(b) shows the calculated C/O atomic ratios for ED samples laser exposed under different atmospheric environments. For the quantification of different carbon species, the C $1s$ core level spectra were fitted using Gaussian Lorentzian sum functions. A representative fitted C $1s$ core level spectrum for ED_Air Exposed (Laser exposed under air environment) is shown in Fig. S6(C). For ED sample at laser unexposed and exposed positions under air environment (ED_Air), the binding energy positions for $sp^3$ carbon were found to be 285.3 eV and 285.0 eV, respectively. For ED sample at laser unexposed and exposed positions under vacuum environment (ED-Vac.), the binding energy positions for $sp^3$ carbon were found to be 285.2 eV and 284.7 eV, respectively. The $sp^2$ carbon for each sample is shifted ~ -1.0 eV w.r.t. $sp^3$ carbon positions. The calculated $sp^2/sp^3$ ratio for different samples has been shown in fig. S6(d).

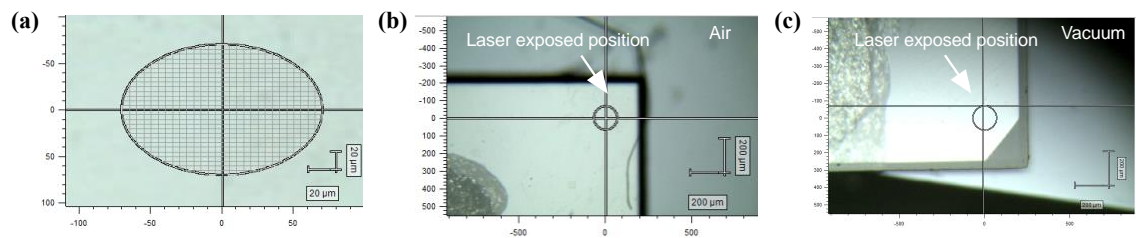

**Figure S5**. Sample preparation for XPS measurements. (a) shows the Raman mapping pattern over diamond. (b) and (c) show the laser exposed region of different samples under air and vacuum respectively.

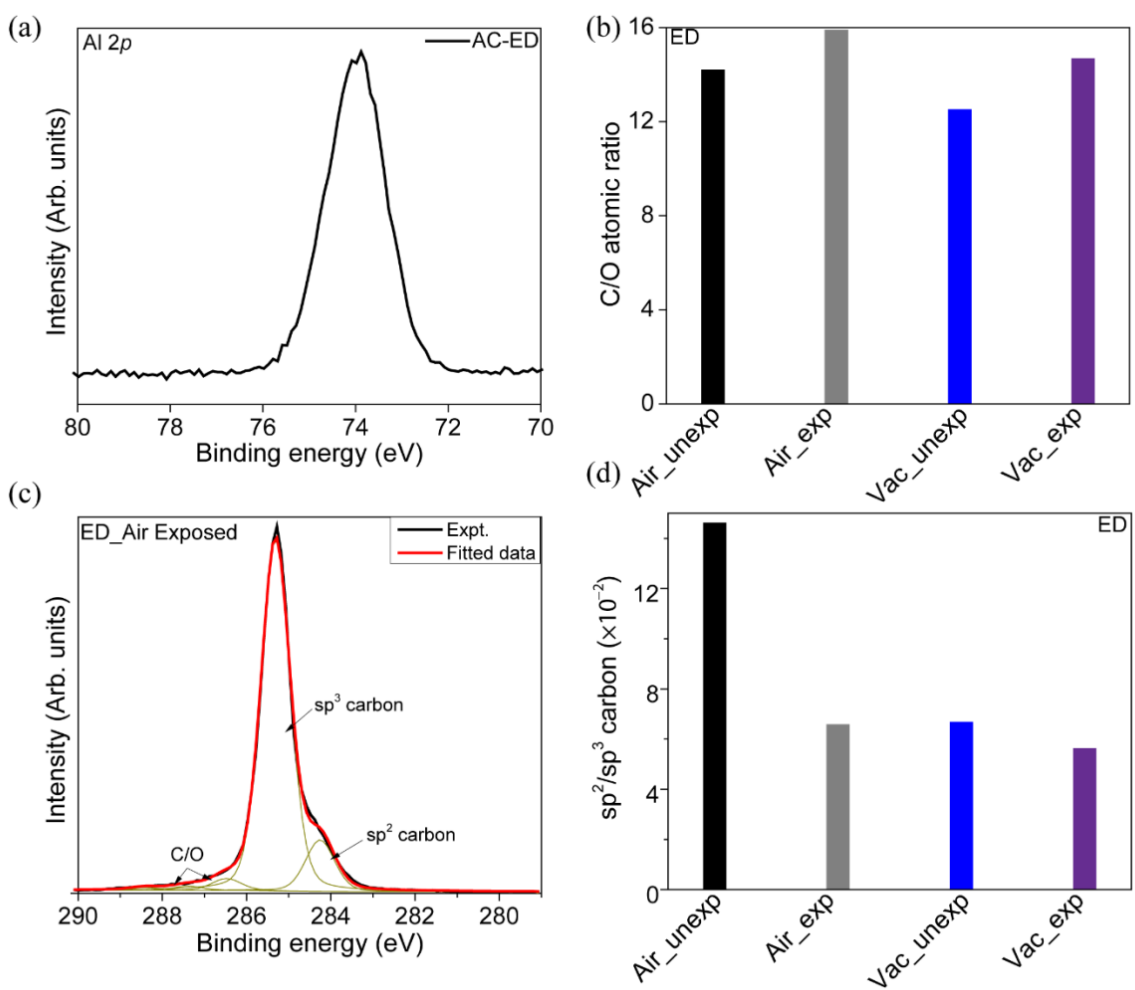

**Figure S6**. (a) Al 2$p$ core level spectrum for AC-ED sample. (b) C/O atomic ratio for EDs after high power laser exposure under different environments. (c) Representative C 1$s$ fitted core level spectrum for sample ED after laser exposure under air environment. (d) The $sp^2/sp^3$ carbon ratio for different ED samples estimated from peak fit analysis of the corresponding C 1$s$ core level spectra.

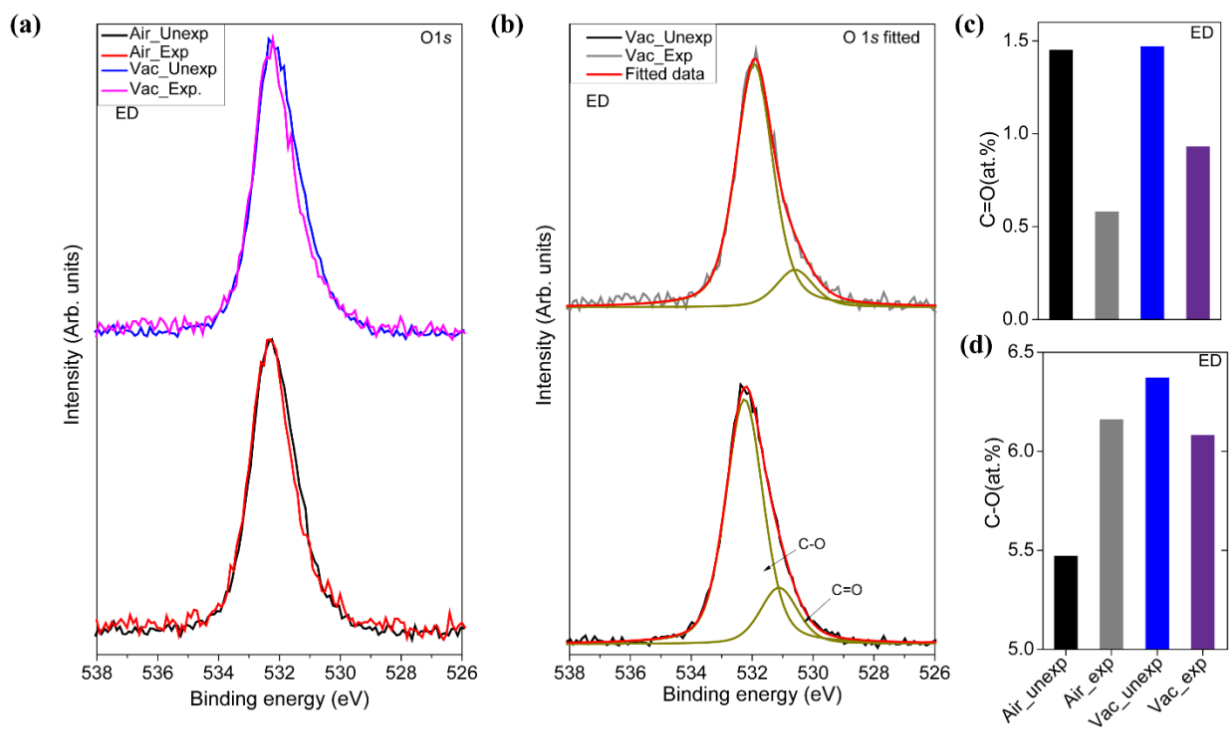

**Figure S7**. (a) O 1*s* core level spectra for EDs after laser exposure under different environments. (b) Peak fit of the O 1*s* core spectrum for ED sample after laser exposure under vacuum environment. (c) and (d) show C-O (%) and C=O (%) respectively for laser exposed EDs under different environments. Here, C-O represents ether (C-O-C) or alcohol (C-O-H) and C=O represents ketone surface functionality.[30, 52]

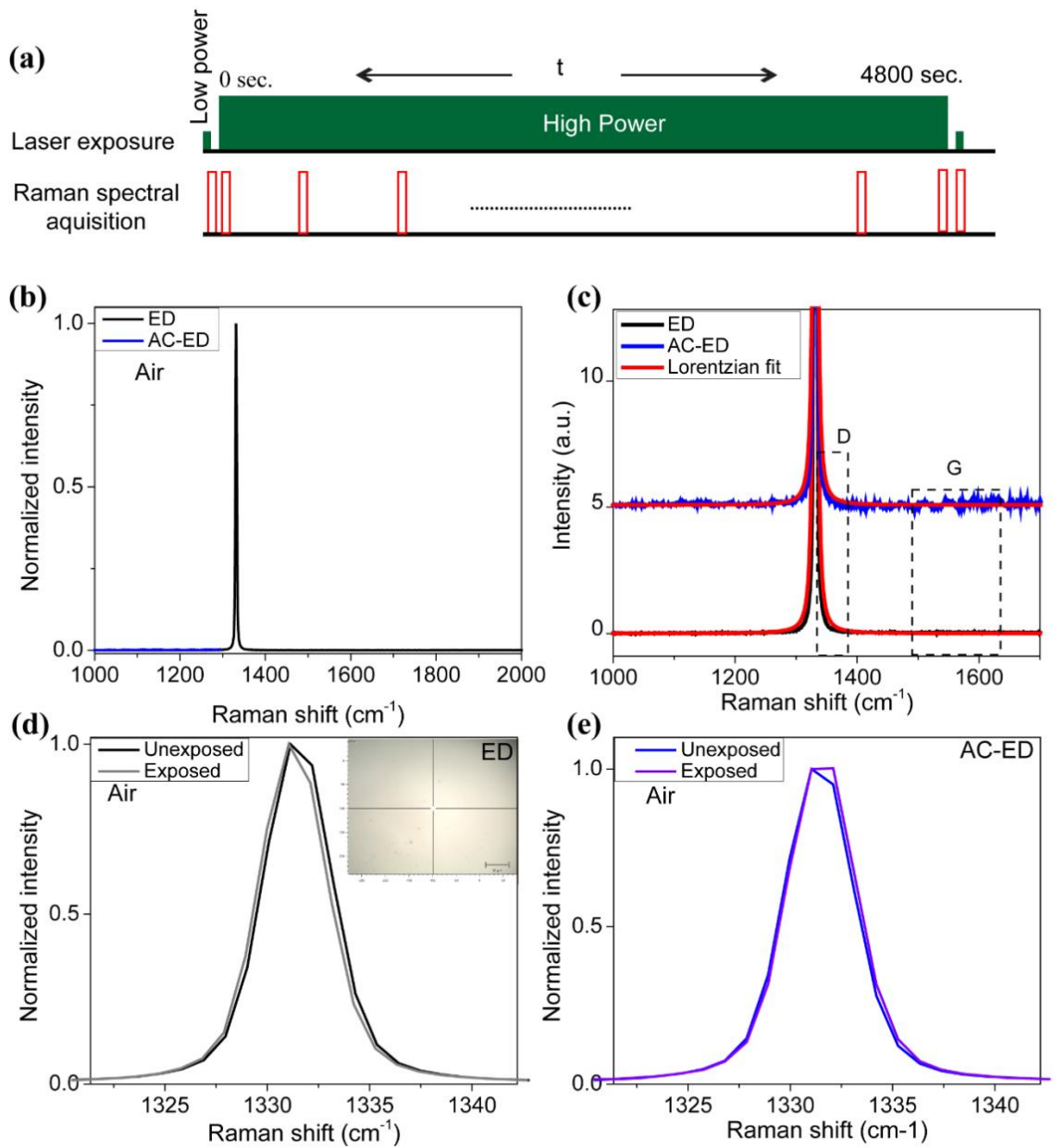

**Figure S8**. a) Experimental scheme for Raman spectroscopy. (b) Raman spectra for ED and AC-ED samples at low excitation laser power and corresponding zoomed D and G band regions. There is no observable feature related with defects (D band) and graphitic carbon (G band) at low laser excitation power. (d) and (e) show the comparison of characteristic diamond Raman features (spectra were acquired at low laser power) before and after high laser power exposure under air environments for ED and AC-ED samples respectively.